# A Compressive Sensing Based Method for Harmonic State Estimation

Saeed Nasiri

*Abstract*— **Power quality monitoring has become a vital need in modern power systems owing to the need for agile operation and troubleshooting scheme. On the other hand, the nature of load in modern power system is changing in many ways. Digital loads, which are mostly relied on power electronic equipment, may distort the quality of power flowing through the network. Moreover, one of the most critical objectives of smart grids is to improve quality of services delivered to customers, alongside with security, reliability and efficiency. To this end, a novel method based on compressive sensing is proposed in this paper to detect the source and the magnitude of the harmonics. The method takes advantages of compressive sensing theory in such a way that a real-time monitoring of harmonic distortion is obtained with a limited number of measurements. The efficacy of the method is checked by means of various simulations on IEEE 118 bus test system. The results show the capabilities of the method in both noisy and noise-free conditions.**

## I. INTRODUCTION

POWER quality based issues have been emerged more than the past, as the use of non-linear digital loads and power electronic based equipment has increased. Moreover, the emerging trends show a rise in injection of intermittent power sources which can cause power quality issues. Among power quality issues, harmonic pollution is identified as one of the most demanding issues to tackle due to the costs they may cause to the network. Nowadays, in a competitive electric market, it is required to monitor harmonic distortion so that both the improvements and the penalties could be fairly decided. Nonetheless, comprehensive information of the sources and amount of harmonics must be in hand of the operator of network to do so [1].

It is worthy of mention that in case of harmonics those equipment that cause the distortion of voltage and currents waves in the network are themselves, sensitive to the harmonic distortion. Harmonic pollution of the power system can result in such problems as spontaneously operation of circuit breakers, overheating of transformers, explosion of the capacitors, malfunction of digital loads and overheating of grounding wires. Electronic based equipment like drives which need the firing angle of thyristors to speed up/down of the electrical motors will also face problems in their operation in presence of harmonics. In general, most of the sensitive power system equipment malfunction if there exist a Total Harmonic Distortion (THD) greater than 5 percent. Harmonic distortion level is of crucial factors to benchmark the quality of power which a utility deliver to its customers. Improving or maintaining the quality of power within an acceptable level considering the emergence of smart grid driven issues is a hard task to achieve. The increase in the penetration of Distributed Energy Resources (DERs) alongside with digital loads can lead the networks to a point that may put the quality of power, especially in harmonic distortion point of view, at risk. The initial step in this regard is the detection and monitoring the condition of the network, instantly. The monitoring of harmonic distortion will result in taking some actions such as designing harmonic filters and installing them in appropriate points of the networks, making regulatory decisions and setting penalties and incentives properly, to tackle the problem.

IEEE Std. 519-1992 requires utilities and consumers to keep their harmonic distortion level within a limit bound. Harmonic State Estimation (HSE) was first introduced in [2] and [3]. HSE aims to identify harmonic sources using a number of monitors installed over the network. The major problem with the paper is that it becomes singular even when the number of monitors is more than that of state variables. In [4] a nonlinear Least Square method and in [5] a Kalman filter based method were proposed. Both these methods suffer from excessive number of monitors needed to install. The former requires 23 monitors for a system with 14 buses and the latter results in installing 54 monitors in the same network.

To compromise the cost of number of monitors, [6] an Artificial Neural Network based method proposed in which the number of real monitors was decreased due to utilizing virtual measurements. The need for a priori information to generate virtual measurements is the main drawback of this method. In [7] a method relied on minimum variance and an underdetermined set of equations was proposed. The major weak point of this method is its need to search all possible solutions to find a correct one, and it requires specifying the location of harmonic sources. [8] uses similar method to [6] and the aforementioned drawback is still true about it. In [9], the location of monitors is obtained using minimization of covariance matrix of real and estimated values. In [10], a method based on Independent Component Analysis (ICA) was proposed to estimate the harmonics. [11] proposed a method to harmonic state estimation in presence of uncertainties. This method takes advantages of Weighted Least Square (WLS) to solve the optimization problem involved in harmonic state estimation.

In this paper, an optimal monitoring method based on Compressive Sensing (CS) is developed. Compressive sensing is an information technology which can offer some useful applications in such area as signal processing, anomaly detection and high dimensional signal representation [12-14].

In power sector, so far, CS has been used in some area of research. In [15] CS is used for the estimation of faults location and current. [16] presents a CS based approach for economic dispatch. Line outage identification based on CS is proposed in [17]. Identification of imbalances using CS is introduced in [18].

By using the proposed method, the location and magnitude of harmonics for each order can be precisely extracted. Additionally, the proposed method relieves the network owner or operator from the need for any priori data of location of harmonic sources. Moreover, the whole network harmonic performance can be assessed for different network operating conditions. Apart from the application of the method for aforementioned reasons, it offers vital advantages for circumstances that the operator faces missing data, false data injection due to cyber-attack or malfunctioning of several monitors.

The rest of this paper is organized as follows. The theoretical background and the concepts of compressive sensing are described in section II. The proposed method to estimate harmonics is discussed in Section III. The numerical results are included in section IV. Finally, section V concludes the main features of the contribution.

## II. COMPRESSIVE SENSING

Compressive sensing is an idea that has gained high interest since its dawn in 2006. The pioneering works on the sparse signal recovery, called as Compressive Sensing (compressed sensing), are introduced in [19-21]. Mathematically, CS seeks to find a unique solution for a linear set of equations with M equations and N unknown quantities in which M<N. In general, such a linear set has infinite solutions, but some features can be assigned to the set through which a unique solution can be found. In fact, the main part of CS is based on defining such conditions that the correct unknown vector in an underdetermined set of equations would be recovered, the number of measurements would be minimized such that the signal is correctly recovered and the recovery procedure would be robust versus the noise of measurements and the sensing matrix. In practice, CS is used in Magnetic Resonance Imaging (MRI) [22]. As the X rays might damage human body while imaging, the lowest possible amounts of X rays is desirable. Therefore, by using CS with a fewer rate of sampling, the image can be recoverable.

As mentioned above, CS needs the signal (unknown vector) and sensing matrix (measurement matrix) to contain some features. The main feature that a signal must have is the sparsity, meaning that in a signal with n number of quantities, k number of quantities are nonzero (in which k<<n) [23]. These kinds of signals are called k-sparse signals. It is worth mentioning that the signals may be non-sparse in a space, and needed to be mapped to another space to be sparse. For instance, a sinusoidal signal is non-sparse (having many nonzero quantities within a period) in the time space and is perfectly sparse after mapping to the Fourier space (with two nonzero quantities represented by Dirac function). Thus, let's

form an underdetermined set of equation as

$$\mathbf{Y}_{m \times 1} = \underbrace{\mathbf{\Phi}_{m \times n} \mathbf{\Psi}_{n \times n}}_{\mathbf{H}_{m \times n}} \mathbf{X}_{n \times 1} \qquad (1)$$

where $\mathbf{Y}_{m \times 1}$ is the measurement vector, $\mathbf{\Phi}_{m \times n}$ is the sensing matrix, $\mathbf{\Psi}_{n \times n}$ is the transformation matrix, $\mathbf{H}_{m \times n}$ is the measurement matrix and $\mathbf{X}_{n \times 1}$ is the unknown sparse signal.

To recover unique correct signal, in addition to the sparsity, other conditions must exist as follows.

**Theorem 1)** All quantities in the k-sparse signal $\mathbf{X}_{n \times 1}$ are uniquely recoverable if and only if the Spark (**H**)>2k.

**Remark 1)** The minimum number of columns of **H** that is linearly dependent is defined as Spark (**H**). Hence, 2 ≤Spark (**H**)≤ Rank(**H**)+1. A proof for theorem 1 is provided in the appendix.

Having Spark condition satisfied, an exhaustive search for the sparsest vector is able to find a unique solution. Another important consequence is that there must be as least 2k number of measurements (m>2k) in order that a unique signal can be recovered. All the above can be mathematically described by an optimization model as

$$min \ \left\| \mathbf{X} \right\|_0 \quad s.t. \quad \mathbf{Y} = \mathbf{H}\mathbf{X} \qquad (2)$$

where, $\left\| \mathbf{X} \right\|_0$ is $L_0$-norm indicating the number of nonzero quantities of vector **X**.

Note that $\left\| . \right\|_0$ is not convex and differentiable meaning that it is computationally Non-Polynomial and thus not applicable when a large number of quantities is involved. In addition, the output of $\left\| \mathbf{X} \right\|_0$ is highly non-linear so that a little variance with zero will result in a jump of $\left\| \mathbf{X} \right\|_0$ by one unit. This apparently shows the sensitivity to the noise. To tackle the noticed issues, an $L_1$-norm ($\left\| . \right\|_1$) can be used instead of $L_0$-norm as an indicator for sparseness as (3) [21].

$$min \ \left\| \mathbf{X} \right\|_1 \quad s.t. \quad \mathbf{Y} = \mathbf{H}\mathbf{X} \qquad (3)$$

In which, $\left\| \mathbf{X} \right\|_1 \triangleq \sum_{i=1}^{n} \left| x_i \right|$ .

Moreover, an $L_1$-norm ($\left\| \mathbf{X} \right\|_1$) is a continuous function of **X** and more robust against the noise so that substituting an $L_0$-norm with an $L_1$-norm converts the optimization model to the one which is convex, and there are efficient methods to solve it [24]. Indeed, it can be shown that the optimization model (4) is an LP problem [25]. Fig.1 depicts different norms capability to reach a correct optimum solution. As illustrated, an $L_1$-minimization can reach a unique solution as an $L_0$ does. However, as it can be intuitively conceived, more restrictions may be required to impose to make sure that the solution obtained is the same as the one reachable by an $L_0$. As such, the norms with higher orders than the $L_1$ (e.g. $L_2$) must have conditions more rigorous to get a correct unique solution [26].

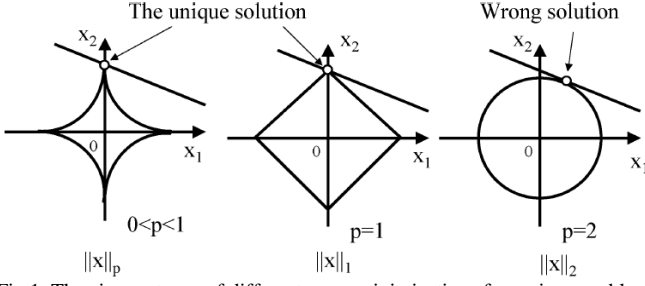

Fig.1. The circumstance of different norm minimizations for a given problem: $\min |x_1|^p + |x_2|^p$ s.t $\mathbf{H_1X_1 + H_2X_2 = c}$.

Using an $L_1$-norm minimization for the signal recovery needs some further criteria assuring that the recovery can be correctly accomplished. In the following, these criteria are introduced.

**Remark 2)** The maximum correlation between two columns of matrix $\mathbf{H}$ is the coherence coefficient of matrix $\mathbf{H}$ and is defined as

$$\mu\left(\mathbf{H}\right) = \max_{i \neq j} \left| \frac{\left\langle \mathbf{H}_i, \mathbf{H}_j \right\rangle}{\left\| \mathbf{H}_i \right\|_2 \left\| \mathbf{H}_j \right\|_2} \right| \quad (4)$$

where $\mathbf{H}_i$ and $\mathbf{H}_j$ are the $i$th and $j$th columns of matrix $\mathbf{H}$.

Though the Null Space Property (NSP) and the Restricted Isometry Property (RIP) are two well-known criteria in terms of signal recovery by compressive sensing, as they are NP, the coherence criterion is widely utilized. The NSP and RIP are presented in the appendix of the paper. Note that, in these two criteria, all the k-sparse vectors must be checked, while with the coherence coefficient only the largest correlation between columns is to be computed. For more details about NSP and RIP, interested readers are referred to [26].

### III. THE METHODOLOGY

The previous section provided some knowledge for compressive sensing as is required for the proposed method of harmonic state estimation. In this section, the proposed method is developed so that the sparsity of voltage sag signal is perceived. Then a sensing matrix suitable for recovering harmonic signal is formed by taking advantage of the coherence condition. Finally, the harmonic signal recovery models are developed.

#### A. Conceptualization

A harmonic estimation problem can be modeled as

$$\mathbf{Y}^h = \mathbf{H}^h\mathbf{X}^h + \mathbf{e}^h$$
$$\begin{pmatrix} \mathbf{Y}_R^h \\ \mathbf{Y}_I^h \end{pmatrix} = \begin{pmatrix} \mathbf{H}_R^h & -\mathbf{H}_I^h \\ \mathbf{H}_I^h & \mathbf{H}_R^h \end{pmatrix} \begin{pmatrix} \mathbf{X}_R^h \\ \mathbf{X}_I^h \end{pmatrix} + \begin{pmatrix} \mathbf{e}_R^h \\ \mathbf{e}_I^h \end{pmatrix} \quad (5)$$

where $\mathbf{Y}$ is the measurement vector composed of several buses voltages and lines currents corresponding to each order of harmonic. $\mathbf{H}$ is a sensing matrix relating the signal to the measurements. $\mathbf{X}$ denotes the signal to be recovered which here is the harmonic current injection at buses and $\mathbf{e}$ is the error between the measurement vector and the sensing matrix multiplied by the signal. Note that this formulation can be separated into real and imaginary parts as shown.

By highlighting two vital features of the compressive sensing once again, a basis will be constructed for a better understanding of the proposed method. The first is the sparsity that either inherently exists or can be created through mapping to another space. The second is the construction of such a measurement matrix that the conditions as previously defined would be satisfied. In case of our purpose, the sparsity does exist inherently by considering the fact that in a power system, the sources injecting the harmonics into the network are sparse. Meaning that, a small number of nodes out of total nodes injects considerable harmonics, simultaneously. Therefore, assuming this the harmonic current injection for each order can be sparse. Thus, a k-sparse signal of harmonic injection will result in 2k number of monitor to install, according to theorem 1. Finally, the following optimization model for a network with $N_b$ number of buses can be modeled as

$$\begin{aligned} min \quad & \left\| \mathbf{X}^h \right\|_1 \\ s.t. \quad & \mathbf{Y}^h = \mathbf{H}^h\mathbf{X}^h \end{aligned} \quad (6)$$

where, $\mathbf{X}^h$ is ($N_b \times 1$) harmonic current injection signal for order h. $\mathbf{H}^h$ is the ($N_m \times N_b$) sensing matrices for harmonic order h ($N_m$ is the number of measurements) which will be formed according to the proposed algorithm in sub-section B. $\mathbf{Y}^h$ are the ($N_m \times 1$) measurement vectors consisting of harmonic instant voltages and currents of given harmonic order at limited buses or lines.

#### B. Sensing matrix design

In CS, the sensing matrix has such significant impact that even with a perfect sparse signal and efficient methods for signal recovery, the signal cannot be correctly recovered without a well-formed sensing matrix. In this paper, in order to form the sensing matrix for harmonic monitoring, an algorithm is proposed as follows.

I. Initially, all the candidate locations and the parameters to be measured must be specified. Knowing that the measurement parameters are the voltages at buses and the lines currents, it can be shown that the dimension of the candidate matrix is ($N_b + N_l$)$\times N_b$, in which $N_l$ is the number of the lines. (7)- (9) are used to relate the elements of the candidate matrix to the measurement vector and to form the candidate sensing matrix along with the corresponding measurement vector.

$$V_i^h = \sum_{n=1}^{N_b} Z_{i,n}^h I_n^h \quad (7)$$

$$I_{ij}^h = V_i^h - V_j^h \; y_{ij}^h + y_{ij,sh}^h V_i^h \quad (8)$$

$$I_{ij}^h = y_{ij}^h \left( \sum_{n=1}^{N_b} Z_{i,n}^h I_n^h - \sum_{n=1}^{N_b} Z_{j,n}^h I_n^h \right) + y_{ij,sh}^h \left( \sum_{n=1}^{N_b} Z_{i,n}^h I_n^h \right) \quad (9)$$

in which i, j and h are the indices for the sending side, the receiving side and the order, respectively. V and I denote the voltages and currents injections. $\mathbf{Z}$ is the impedance matrix and y is the index for the admittance imbedded in a pi model of the line corresponding. Note that passive and active elements of power systems are modeled based on [71,72].

II. Based on theorem 1, a given number of rows should be chosen to form the sensing matrix and, in turn, to know that which points must be measured by installing the monitors. According to theorem 1, the number of the sensing matrix rows has to be, at least, two times greater than the sparsity of the signal. Note that this may not be necessarily equal to the number of the monitors. Hence, once the candidate matrix is completely formed, a sensing matrix composed of a limited pre-defined number of rows will be formed by means of an optimization model as (10). This model indeed minimizes the coherence of sensing matrices related to each order. It is worthy of mention that the Genetic Algorithm is utilized here to reach an optimal solution. The desired number of monitors to be installed is the input of the algorithm. Then, lots of matrices are examined as potential solutions for the sensing matrix by the objective function of (10). Finally, the best solution is chosen as the optimal sensing matrix, and the locations for installing the monitors are extracted.

$$min \; \mu \quad \mathbf{H} \; = \sum_h \frac{1}{h} \left\| \mathbf{H}^{h*} \mathbf{H}^h \, \text{-} \, \mathbf{I} \right\|_\infty \qquad (10)$$
$$s.t. \; spark(\mathbf{H}^h) \geq 2k$$

### C. Signal recovery

There are generally two approaches to reconstruct the sparse signals, including convex based optimization and the greedy methods. The earlier gives more accurate outcome but inapplicable in large dimension cases due to excessive time consumed. While the greedy methods are developed to eliminate such a drawback, the probability of reconstructing of the correct signal might be reduced.

The convex based approaches, indeed, convert the norm optimization models to an LP one or similar models such as quadratic programing which can be solved by the analytic methods. The methods using an LP model to solve an L1-norm optimization problem are so called BP (Basis Pursuit) family methods [27-30]. Note that such optimization model as (3) can be formulated by an LP model as (11) - (13).

$$\begin{cases} \underset{\mathbf{x}}{minimize} & \mathbf{C}^T \mathbf{X}^h \\ s.t. & \begin{cases} \mathbf{A}\mathbf{X}^h \leq \mathbf{b} \\ \mathbf{X}^h \geq \mathbf{0} \end{cases} \end{cases} \qquad (11)$$

$$\mathbf{X}^h = \mathbf{X}^h_+ \, \text{-} \, \mathbf{X}^h_- \Rightarrow \left\| \mathbf{X}^h \right\|_1 = 1,1,\cdots,1 _{1 \times 2n} \begin{pmatrix} \mathbf{X}^h_+ \\ \mathbf{X}^h_- \end{pmatrix}_{2n \times 1} \qquad (12)$$

$$\mathbf{H}^h \mathbf{X}^h = \mathbf{Y}^h \Rightarrow \mathbf{H}^h \, \text{-} \, \mathbf{H}^h _{m \times 2n} \begin{pmatrix} \mathbf{X}^h_+ \\ \mathbf{X}^h_- \end{pmatrix}_{2n \times 1} = \mathbf{Y}^h$$
$$\begin{pmatrix} \mathbf{H}^h & \text{-} \mathbf{H}^h \\ \text{-} \mathbf{H}^h & \mathbf{H}^h \end{pmatrix}_{2m \times 2n} \begin{pmatrix} \mathbf{X}^h_+ \\ \mathbf{X}^h_- \end{pmatrix}_{2n \times 1} \leq \begin{pmatrix} \mathbf{Y}^h \\ \text{-} \mathbf{Y}^h \end{pmatrix}_{2n \times 1} \qquad (13)$$

where, $\mathbf{X}_+$ is a vector with the same length as $\mathbf{X}$ in which only positive quantities of $\mathbf{X}$ are included and the rest is zero. Similarly, $\mathbf{X}_-$ is for the negative quantities with the exception that this includes absolute values without their signs. An L1-norm optimization will, therefore, be converted to an LP problem readily solvable by Simplex or point interior algorithms. As well, when a noisy case is supposed, (14) and (15) can be used and a similar procedure can be applied to convert them to an LP model.

$$\begin{cases} \underset{x}{min} & \left\| \mathbf{X}^h \right\|_1 \\ s.t. & \left\| \mathbf{Y}^h \, \text{-} \, \mathbf{H}^h \mathbf{X}^h \right\|_1 \leq \varepsilon \end{cases} \qquad (14)$$

$$\begin{cases} \underset{x}{min} & \left\| \mathbf{X}^h \right\|_1 \\ s.t. & \left\| \mathbf{Y}^h \, \text{-} \, \mathbf{H}^h \mathbf{X}^h \right\|_\infty \leq \varepsilon \end{cases} \qquad (15)$$

in which, $\varepsilon$ is an allowable error in a noisy condition. It is worth noting that in (14), all the errors are considered to be constrained but in (15), only the maximum error among all is assumed to be effective.

There are other well-known algorithms such as BPDN (Basis Pursuit DeNoising), LASSO (Least Absolute Shrinkage and Selection Operator) and DS (Dantzig Selector) to deal with an L1-norm minimization by means of convex based methods. The formulation of the optimization models based on the above algorithms is represented by (16)-(18), respectively.

$$\begin{cases} \underset{x}{min} & \left\| \mathbf{X}^h \right\|_1 \\ s.t. & \left\| \mathbf{Y}^h \, \text{-} \, \mathbf{H}^h \mathbf{X}^h \right\|_2 \leq \varepsilon \end{cases} \qquad (16)$$

$$min \quad \left\| \mathbf{Y}^h \, \text{-} \, \mathbf{H}^h \mathbf{X}^h \right\|_2^2 + \lambda \left\| \mathbf{X}^h \right\|_1 \qquad (17)$$

$$\begin{cases} \underset{x}{min} & \left\| \mathbf{X}^h \right\|_1 \\ s.t. & \left\| \mathbf{H}^{h\mathbf{T}} \, \mathbf{Y}^h \, \text{-} \, \mathbf{H}^h \mathbf{X}^h \right\|_\infty \leq \varepsilon \end{cases} \qquad (18)$$

In recent years, many greedy methods have been proposed, of which MP (Matching Pursuit), OMP (Orthogonal Matching Pursuit), IHT (Iterative Hard Thresholding) and CoSaMP (Compressive Sampling Matching Pursuit) can be mentioned [31-35].

Thus, (6) can be efficiently solved by these tools and then the harmonic current injections can be obtained. In the final stage, using (19), the voltage magnitude of each bus can be found for harmonic order h.

$$\mathbf{V}^h = \mathbf{Z}_{bus}^h \mathbf{X}^h \qquad (19)$$

### IV. THE NUMERICAL STUDY AND ANALYSIS

In this section, the IEEE 118-bus test system is presented to analyze the proposed method. The IEEE 118-bus is composed of 35 generating stations, 177 lines connecting 118 buses and nine transformers. The data for the system is provided in [54]. As mentioned before, the sparsity of harmonic distortion over the network must be defined. Here, we assume that 30 nodes of the IEEE 188-bus, at most, can pollute the network at the same time. Therefore, the spark of sensing matrix must be greater than 60 which will lead to install 60 monitors.

### A. Noise-free monitors, perfect sparsity and accurate model

As mentioned, for k simultaneous harmonic injections with order h, and theoretically, there must be at least 2k number of monitors. The sensing matrix and the locations of the monitors are determined through solving (10). It is worth noting that the harmonic orders to be considered are odd orders up to 23rd harmonic order. However, as the less the order of harmonics, the more impact they have, the significance of each order is assigned with its order inversely. Nonetheless, note that some order of harmonics could cause more damage, as its frequency stimulates one of the natural frequencies of the system.

To test the capability of the method a simulation is conducted as follows. At first, 30 buses are chosen randomly using a uniform distribution among 118 buses. Then, the injection harmonic currents are determined using a normal distribution with an enough small mean and standard deviation related to each harmonic order. In the next step, using (7)-(9), the actual voltages and currents are computed. These values will be used as measurement vector corresponding to sensing matrix and monitor installation points. Finally, the optimization model (6) is solved by utilizing the CVX [36] optimization toolbox.

Simulation results show that all the chosen locations are lines currents while none of voltage channels of monitors are used. Note that it is the sensing matrix formed by (10) that determines how the monitors must be deployed. The results for all buses and different order of harmonics are illustrated in Fig.2, indicating the actual magnitude of current injection and voltages at buses and the recovered ones by the proposed method.

It can be seen that the magnitudes of harmonic currents and voltages are recovered so accurately that the absolute difference between the actual and recovered signals is close to zero at most buses especially in case of harmonic voltages. Note that in some harmonic orders, the magnitudes of voltages are considerable owing to the effect in which the frequency of harmonic resulted in the resonance in some branches of the network. This happening depends on the location of harmonic sources, apart from the impedance model of network. The results obtained apparently show the efficacy of the method even in a large scale power system when the monitors are noise-free, the impedance model of network is accurate enough and the harmonic injection signal is perfectly sparse (i.e the vector of harmonic injection has some non-zero elements and the rest elements are zero).

However, in real power system such assumptions may not exist due to some reasons. Firstly, the data gathering process face many risks such as loss of data when transferring them to the control center and a level of inaccuracy due to the sending and receiving data in a noisy environment. Secondly, as our method relies on impedance model of the network, it is possible that the impedance value of different elements of the network may not be accurate enough at least in case of some harmonic orders. Finally, the perfect sparsity assumption may not be a realistic hypothesis supported strongly by the facts of power systems. In fact, in terms of harmonic injections, some buses inject a considerable amount of distorted current into the network and the others may only inject negligible amount of distorted current into the network. In the next, we investigate the capability of the method with respect to the issues just mentioned.

### B. Noisy monitors, imperfect sparsity and inaccurate model

To discover the proposed method aptitude in such condition that the monitors are noisy, the impedance model of the network is inaccurate and there exist some non-zero harmonic injections but small enough to be neglected, an extreme trial is conducted as follows.

To take the noise of measurement into account a white noise using a normal distribution with a mean of zero and a standard deviation of 5% is added to the values which are used as measurement vector in the model. To model the inaccuracy of impedance model of the network, the values sensing matrix is distorted by adding a random value to each element of the matrix utilizing a normal distribution with the same parameters. As such, to make the harmonic injection signal imperfectly sparse, all buses, other than those with considerable harmonic injection which are determined randomly, are assumed to have harmonic injections so that their magnitudes are considered to be a random number with a mean of zero and a standard deviation of 5% of the greatest harmonic source.

Then, the simulation conducted in previous section was repeated but model (20) is employed setting $\varepsilon$ as 0.01 to recover the signal.

$$\begin{array}{ll} \underset{x}{minimize} & \left\| \mathbf{X}^h \right\|_\infty \\ s.t. & \left\| \mathbf{Y}^h - \mathbf{H}^h \mathbf{X}^h \right\|_2 \leq \varepsilon \end{array} \qquad (20)$$

It is worth noting that many simulations were conducted while only one of the aforementioned origins of noise was modeled. The results showed that the worst case is the case in which all three sources of uncertainty (i.e monitors noise, inaccurate model and imperfect sparsity) are presented at the same time. Therefore, the combination of the explained procedure for noise modeling was used. Fig.3 depicts the results of the method proposed for harmonic state estimation when an acute case is simulated. As shown, the method is done well even with noisy monitors, inaccurate impedance model and imperfect sparsity so that the difference between the actual and recovered signal of harmonic injections remains below 0.1 P.U. for current injections, and close to zero for harmonic voltages. Note that, although the exact values of harmonic injections at some buses are not recovered precisely, all the locations of harmonic injections are detected correctly. However, it should be pointed out that an extreme case is considered to benchmark the efficacy of the method for estimating harmonics. Apparently, the method is capable of recovering the injection signal with an acceptable level of accuracy if a normal realistic condition would be assumed.

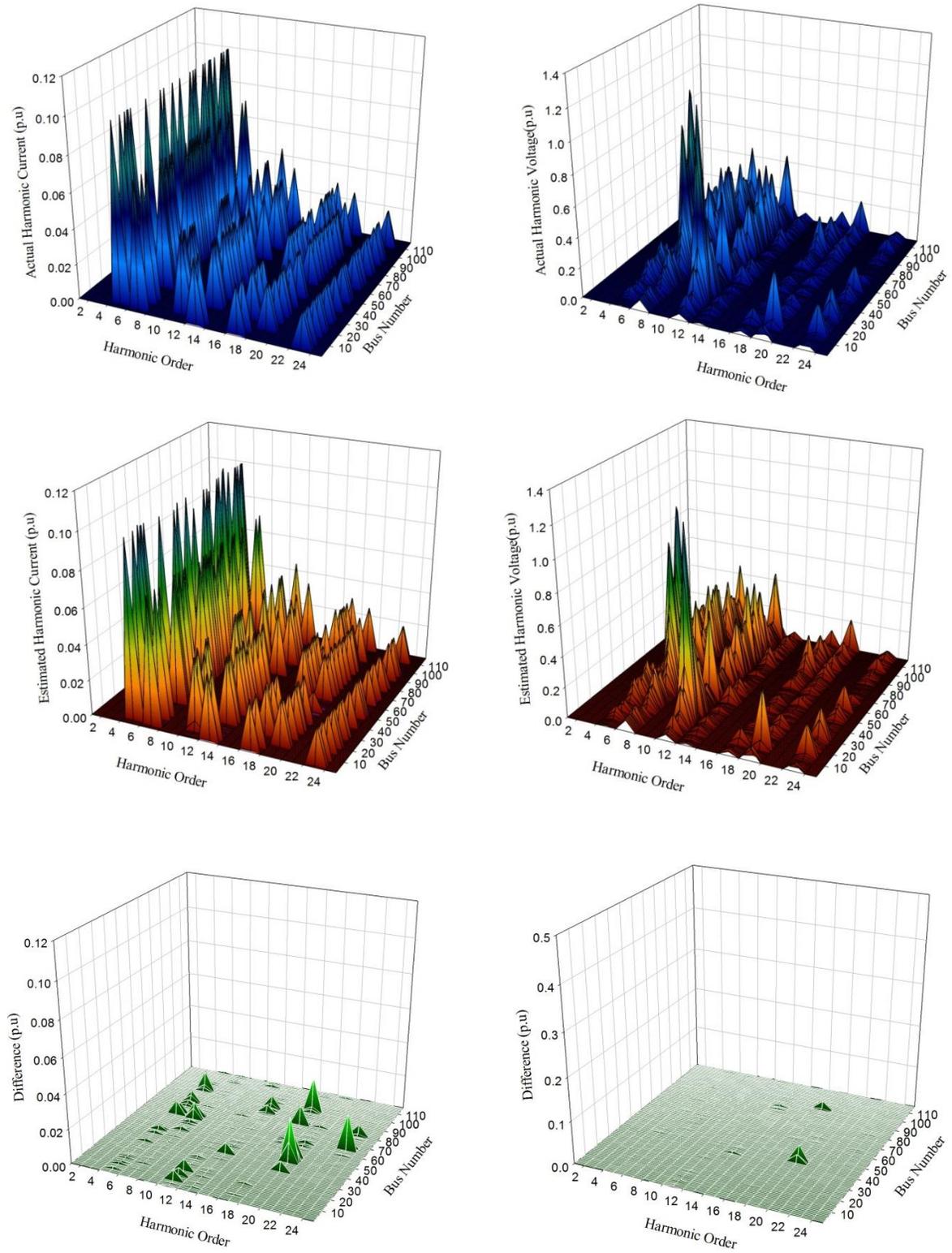

Fig.3. The actual and recovered voltage magnitudes and harmonic currents for different harmonic orders and all buses with noise-free condition and perfect sparsity

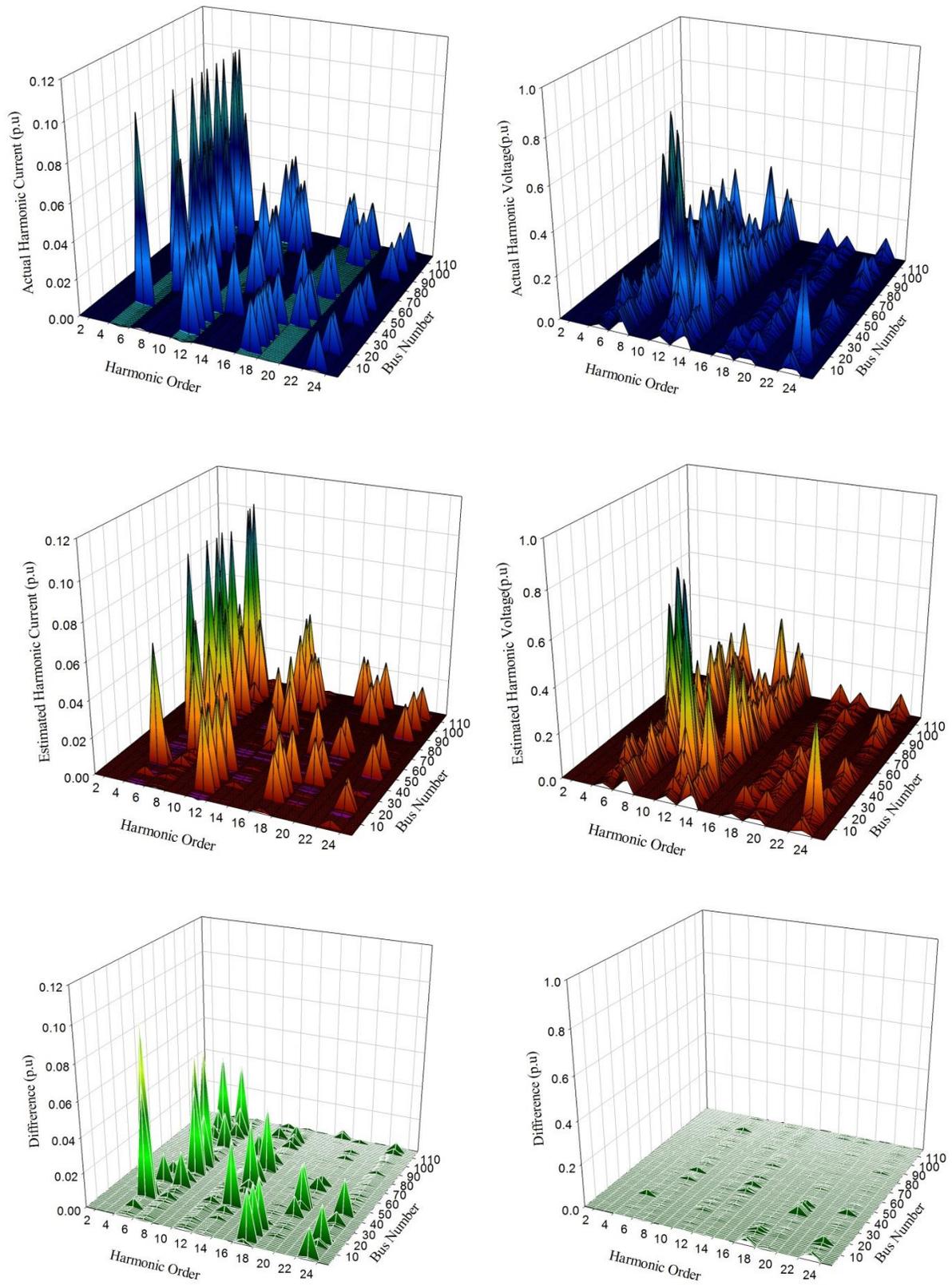

Fig.3. The actual and recovered voltage magnitudes and harmonic currents for different harmonic orders and all buses with noisy condition and imperfect sparsity

## V. CONCLUSIONS

This paper has developed a comprehensive approach to monitor harmonic distortions, as one of the the most important power quality disturbances. Though the existing methods for harmonic state estimation try to cope with the number of monitors, the method proposed in this paper addresses the observability issues using a sparse representation of harmonic injections over the network.

The paper has provided a thorough knowledge of the compressive sensing required to the proposed method of monitoring. Furthermore, the efficacy of the method was shown in an extremely noisy condition by means of an acute benchmark with a large scale power system test case.

## VI. APPENDIX

To prove theorem 1, a contradictory argument can be simply used. Suppose that we have the measurement matrix $\mathbf{H}$ satisfying Spark $(\mathbf{H})>2k$ and $\mathbf{X_1}$ and $\mathbf{X_2}$ are two different k-sparse solutions for $\mathbf{Y=HX}$. We can then get

$$\mathbf{HX_1 = HX_2} \quad \rightarrow \quad \mathbf{H(X_1 - X_2) = 0}$$
$$\mathbf{G \equiv X_1 - X_2} \quad \rightarrow \quad [\overrightarrow{\mathbf{H_1}}, \overrightarrow{\mathbf{H_2}}, \cdots, \overrightarrow{\mathbf{H_n}}]\mathbf{G} = \sum H_{i,j} = 0 \tag{21}$$

The final conclusion of (21) contradicts our initial assumption, since $\mathbf{X_1}$ and $\mathbf{X_2}$ are two unique k-sparse vectors and their subtraction is a 2k-sparse vector, not a zero vector.
**Theorem 2)** Matrix $\mathbf{H}$ satisfies the NSP (Null Space Property) criterion with order k and coefficient $C_{NSP}$ if

$$\forall s \in 1, 2, \cdots, n , \overrightarrow{\mathbf{v}} \in null(\mathbf{H}):$$
$$\left\|\overrightarrow{\mathbf{v}_s}\right\|_1 \leq C_{NSP}\left\|\overrightarrow{\mathbf{v}_{s^c}}\right\|_1, |s| = k \tag{22}$$

Indeed, this restricts the sparse vectors such that they do not take a place in a null space of $\mathbf{H}$. The sufficient and necessary condition to recover the sparse signal is given by the NSP condition. Nonetheless, to guarantee the signal reconstruction in the presence of noise, a stricter condition than NSP is required.
**Theorem 3)** Matrix $\mathbf{H}$ satisfies the condition RIP (Restricted Isometry Property) with order k and a constant $\delta_k \in [0,1)$ if

$$(1-\delta_k)\left\|\mathbf{X}\right\|_2^2 \leq \left\|\mathbf{HX}\right\|_2^2 \leq (1+\delta_k)\left\|\mathbf{X}\right\|_2^2 \tag{23}$$

This ensures the length of k-sparse vectors to be approximately fixed. Hence, if matrix $\mathbf{H}$ has the condition RIP with order 2k, the distance between two k-sparse vectors will be kept so safe that the k-sparse vectors can be separable, even with some noise. It can be mathematically written as

$$\mathbf{X_1}, \mathbf{X_2} : k - Sparse \Rightarrow \mathbf{X = X_1 - X_2} : 2k - Sparse$$
$$(1-2\delta_k)\left\|\mathbf{X_1 - X_2}\right\|_2^2 \leq \left\|\mathbf{H(X_1 - X_2)}\right\|_2^2 \leq (1+2\delta_k)\left\|\mathbf{X_1 - X_2}\right\|_2^2 \tag{24}$$